\documentclass[pra,twocolumn,showpacs,amsmath,amssymb]
{revtex4}

\usepackage[dvips]{color}
\usepackage{graphicx}

\begin{document}

\title{Dynamical studies of macroscopic superposition states:
Phase engineering of controlled entangled number states of
Bose-Einstein condensates in multiple wells}

\author{Mary Ann Leung,$^1$ Khan W. Mahmud,$^2$ and William P. Reinhardt$^{1,3}$}
\affiliation{$^1$Department of Chemistry, University of
Washington, Seattle, WA 98195-1700, USA \\
$^2$Department of Physics, University of California,
Davis, California 95616, USA \\
$^3$Department of Physics, University of Washington, Seattle, WA
98195-1560, USA}

\begin{abstract}
We provide a scheme for the generation of entangled number states
of Bose-Einstein condensates in multiple wells with cyclic
pairwise connectivity. The condensate ground state in a multiple
well trap can self-evolve, when phase engineered with specific
initial phase differences between the neighboring wells, to a
macroscopic superposition state with controllable entanglement --
to multiple well generalization of double well NOON states. We
demonstrate through numerical simulations the creation of
entangled states in three and four wells and then explore the
creation of ``larger" entangled states where there are either a
larger number of particles in each well or a larger number of
wells. The type of entanglement produced as the particle numbers,
or interaction strength, increases changes in a novel and
initially unexpected manner.
\end{abstract}


\maketitle

\section{Introduction}

Entanglement, a nonclassical correlation between two or more
physical systems, lies at the heart of the profound difference
between quantum mechanics and a local classical description of the
world~\cite{einstein1935}. Apart from their discussions in the
philosophical and foundational aspects of quantum
mechanics~\cite{bell1987}, entangled states in recent years have
become an essential resource for the emerging field of quantum
information processing~\cite{horodecki2009}. Entangled and
squeezed states hold promise in studies related to quantum
measurement, Heisenberg limited atom interferometry and precision
measurements~\cite{huelga1997,pezze09}, and quantum computing and
quantum communication. Since multi-particle entangled states can
be more useful than the two-particle/two-photon states, there have
been steady attempts toward creating such
states~\cite{monroe2002}. Such states have been created with
several systems -- with five photons~\cite{zhao2004}, eight atoms
in an ion trap~\cite{haffner2005}, ten nuclear spins in a
molecule~\cite{jones2009}, and also with cold atoms in an optical
lattice~\cite{mandel2003, oberthaler2008}. There have been several
proposals to create superposition states with a large number of
particles such as using a tiny mirror~\cite{marshall03} and
microorganisms such as viruses~\cite{cirac10a}.

While the consequence of entanglement for an
Einstein-Podolsky-Rosen (EPR) pair is quantified in Bell's
inequality~\cite{bell1965}, a more striking conflict between
quantum mechanics and local realism is exhibited by three
maximally entangled spins also known as the
Greenberger-Horne-Zeilinger (GHZ) states~\cite{greenberger1990}.
GHZ state of $N$ spins has the form
\begin{equation}|\Psi \rangle =\frac{1}{\sqrt{2}}\left( |{N},0\rangle+|0,{N}\rangle\right)
\label{eqn:ghz}
\end{equation}
This state can also be written in the notation
$|00...0\rangle+|11...1\rangle$ where $|0\rangle$ and $|1\rangle$
are the basis states. The superposition of two macroscopically
distinct states, rather than simply the internal degrees of
freedom, each occupied by all $N$ particles, has been discussed by
Schr\"{o}dinger in the famous cat parable~\cite{sch1935}; partial
realization of such Schr\"{o}dinger's cat states has been obtained
with Josephson junction loops~\cite{friedman00,lloyd00}. That
macroscopic superposition states are highly entangled has been
discussed in Ref.~\cite{morimae10}.

The analog for the GHZ state for ${N}$ particles in two wells is
denoted like Eq.~\ref{eqn:ghz}, $\frac{1}{\sqrt{2}}(|{N},0\rangle$
+ $|0,{N}\rangle)$, and is colloquially referred to as ``NOON"
states. A generalization of this two state model to multi
dimensions has been discussed in Ref.~\cite{cerf2002}. In this
paper, we discuss the generation of macroscopic entangled number
states of a multiwell Bose-Einstein condensates (BEC) of the
approximate form
\begin{eqnarray}
|\Psi \rangle_{ \{ {N}, M \}^{max}}
&=&\frac{1}{\sqrt{M}}( |{N},0,0,\ldots,0\rangle+|0,{N},0,\ldots,0\rangle \nonumber \\
&&+\ldots +|0,0,\ldots,{N} \rangle \label{eqn:extremeCat}
\end{eqnarray}
This is the multiwell generalization of the double well NOON
states where a macroscopic number of particles are simultaneously
in $M$ different locations, with $M > 2$. BEC in optical
lattices~\cite{review} has been a promising research area with
many new observations such as the superfluid to Mott insulator
transition~\cite{greiner2002} and number-squeezed
states~\cite{orzel2001}. Superfluid and Mott insulator states are
the ground states of bosons in an optical lattice, whereas the
multi-positional Schrodinger cat state of type
Eq.~\ref{eqn:extremeCat} is the highest lying excited state. Due
to the coherence properties and versatility of cold atom systems,
it may be an ideal system to create such entangled states.

We show that states approximating the extreme entangled states of
Eq.~\ref{eqn:extremeCat} may be generated in a controlled fashion
by time evolution of appropriately phase imprinted ground states
of a multiwell BEC with periodic boundary conditions for $M=3, 4$,
and $8$. The physical mechanism for creating such states can be
understood from the phase space picture of a double well
BEC~\cite{mahmud2005,mahmudThesis} where it was shown that the
ground state wave packet displaced in phase and put on a
hyperbolic fixed point of its underlying semiclassical phase space
dynamically bifurcates to a macroscopic superposition state, a
highly entangled state. Similarly, for a multi-well BEC, phase
imprinting the ground state moves it to an unstable equilibrium,
and subsequent dynamics creates entangled states. We show that the
choice of initial barrier heights, which determine the extent of
ground state number squeezing, and the rate of barrier ramping can
be used to control the entanglement of the final states.

Based on results obtained for two, three, and four well
configurations, we conjecture a generalized formula, for $M$
wells, for the phase offset between neighboring wells appropriate
for the generation of number entangled states.  Finally, we extend
our analysis to larger systems where there are a larger number of
particles in each well or a larger number of wells.  In these
cases, we find surprising results indicating the formation of a
new type of number entangled state in four wells.  We study the
impact of increasing the number of particles or the interaction
strength on the resultant entangled state.

The paper is organized as follows. In Sec II, we introduce our
model and discuss the numerical methods. In Sec III, we review
previously published~\cite{mahmud2005} results for the generation
of NOON like entangled states in a double well. This is done in
order to motivate and generalize the double well results to
multiple wells. In Sec IV, we present our study of three and four
wells showing the phase engineering method for creating entangled
states. We focus on small number of particles, mainly showing that
the method works for multiple wells, and that we can control the
final state by controlling the initial barrier height and barrier
ramps. In Sec V, we study this for larger number of particles in
four and eight wells, and find that a new type of entangled number
state is also created in the process. Finally, we summarize our
results in conclusion in Sec VI.

\section{Methods}
%
\begin{figure}[ht]
\begin{center}
\includegraphics[width=0.45\textwidth,angle=0]{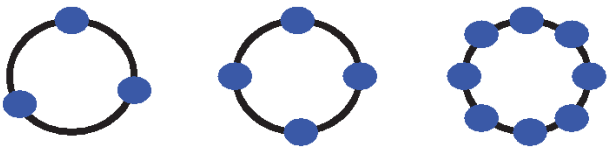}
\end{center}
\caption{\label{fig:multiwellconfig} Shown are multi-well
configurations for three, four, and eight wells with periodic
boundary conditions in a one dimensional circular array.}
\end{figure}

The physical configuration we study assumes multiple wells
connected in a circular array, which has been realized
experimentally by~\cite{amico2005}, and is illustrated in
Fig.~\ref{fig:multiwellconfig}.

We approximate the physics of a BEC in a multiwell potential by
the Bose-Hubbard (BH) model~\cite{fisher1989,jaksch1998}. Thus
\begin{eqnarray}
\hat{H} &=&-J\sum_{i}(a_{i}^{\dagger }a_{i+1}+a_{i+1}^{\dagger
}a_{i})+\sum_{i}\epsilon _{i}\hat{n_{i}}  \nonumber \\
&&+\frac{1}{2}U\sum_{i}\hat{n_{i}}(\hat{n_{i}}-1)
\label{eqn:BHhamil}
\end{eqnarray}
where $\hat{n_{i}}=a_{i}^{\dagger }a_{i}$ is the number operator,
$J$ is the nearest neighbor tunneling term, $U$ is the on-site
energy, and $\epsilon _{i}$ is the energy offset of the $i$th
lattice. To simplify a theoretical study, we make a one parameter
approximation of the tunneling and interaction strength:
$U/J=1/e^{-\alpha(t)}$; and for the symmetric wells explored here,
$\epsilon _{i}=0$. Because of the tunability in optical lattices,
$U$ and $J$ can be changed in time with varying lattice depth, and
we take $\alpha(t)$ as a function of time $t$. $\alpha(t)$ is a
dimensionless parameter that can be mapped onto the barrier
height. This parametrization allows a simple study of continuous
change of barrier height through the variation of a single
parameter $\alpha(t)$. For example, for a lattice made of red
detuned laser with $\lambda =985$ nm and for $^{23}$Na, a barrier
height $15E_{R}$ gives $U=0.15E_{R}$ and
$J=0.07E_{R}$~\cite{jaksch1998} where
$E_{R}=\frac{\hbar^{2}k^{2}}{2m}$ is the recoil energy from
absorption of a photon; these experimental parameters then
correspond to $\alpha =2.14$.

Our numerical studies focus on three size regimes: I) a small
number of particles in a small number of wells, II) a large number
of particles in small number of wells, similar to the experiments
of~\cite{orzel2001}, and III) a small number of particles in a
larger number of wells, such as the experiments
of~\cite{greiner2002}. For ${N}$ identical bosons in $M$ wells the
number of Hilbert space dimension is
$D=\frac{({N}+M-1)!}{{N}!(M-1)!}$. The latter two cases result in
large $D$ and necessitate the use of parallel processing
techniques. The largest systems we have investigated at the time
of writing is 512 particles in four wells with 202 million nonzero
entries, $D$=22,632,705 and 24 particles in eight wells,
$D$=2,629,575 with 3.5 million nonzero entries. Details of our
parallel implementation of the Bose Hubbard model can be found in
\cite{leung2007}.

The main equation we solve is the time dependent Schr\"{o}dinger
equation (TDSE):
\begin{equation}
\label{eqn:discreteSch} \sum_k[H]_{jk}c_k(t)=i \hbar \frac {d
c_j(t)}{dt}
\end{equation}
where $H$ is the discretized BH Hamiltonian in the Fock state
basis $|m\rangle$. The solution of the time dependent
Schr\"{o}dinger equation in the full Fock space is then of the
form
\begin{equation}
\label{eqn:fockgen} |\Psi(t)
\rangle=\sum_{m=0}^{D-1}{c_m(t)|m\rangle}
\end{equation}
%

\section{Entangled number states of the BEC in two wells}
%
\begin{figure}[ht]
\includegraphics[width=0.45\textwidth,angle=0]{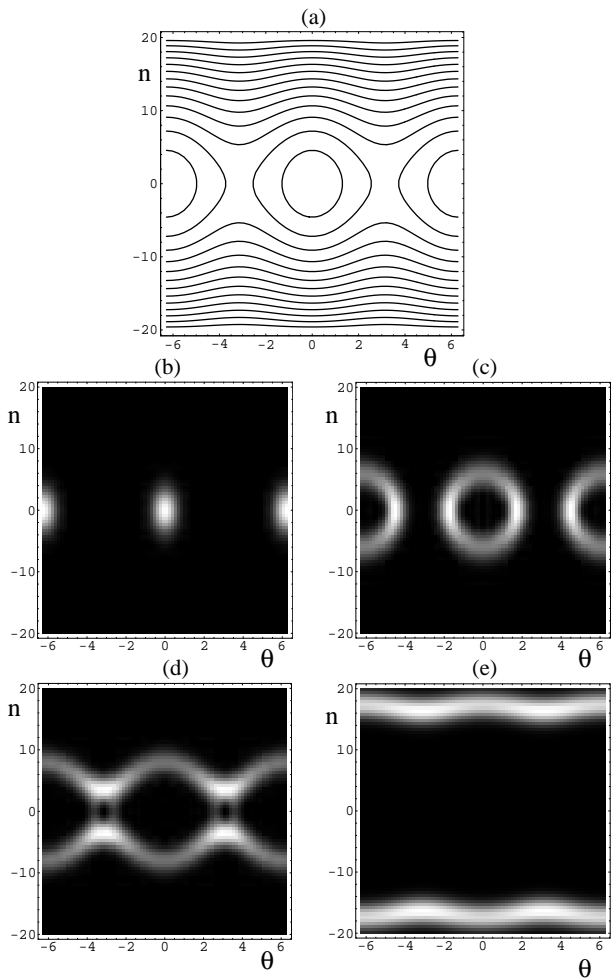}
\caption{\label{fig:phasespace} Comparison of the classical
nonrigid physical pendulum phase space with quantum phase space
shown with the Husimi probability distributions for double well
eigenstates. Shown are (a) classical energy contour. Husimi
projections for (b) ground state which is minimum uncertainty
wavepacket centered at the origin, (c) 6th state, (d) 12th state,
and (e) 35th state which is analogous to a superposition of
classical pendulum rotor motions in two opposite directions. The
analogy between the double well BEC and physical pendulum points
to a way to the creation of macroscopic superposition states by
displacing the ground state to the unstable equilibrium points in
the classical phase space.}
\end{figure}

In order to generalize the creation of entangled states in
multiple wells, we briefly review here the physical principles
behind creating such states in a double well which was described
in Ref.~\cite{mahmud2005,mahmudThesis}. The physics of creating
such states in a double well as well as its extension to multiple
wells as presented here can be understood in terms of the
underlying classical phase space dynamics.

The most general state vector in a double well is a superposition
of all the number states
\begin{equation}
|\Psi\rangle=\sum_{n_L=0}^{N}c^{(i)}_{n_L}|n_L,N-n_L\rangle
\label{eqn:basis}
\end{equation}
where $n_L$ is the number of particles in the left well, and $N$
the total number of particles.  Finding the eigenvalues and
eigenvectors of the Bose-Hubbard hamiltonian for two wells can be
easily accomplished by diagonalizing a $(N+1) \times (N+1)$
tridiagonal matrix, getting the coefficients $c^{(i)}_{n_L}$.

After we implement the cat state generation method, states of the
following form can be generated,
\begin{equation}
|\Psi\rangle=\frac{1}{\sqrt{2}}\left(|N-n_L,n_L\rangle+|n_L,N-n_L\rangle
\right)
\end{equation}
When $n_L=0$ or $N$, it becomes an extreme superposition state
\begin{equation}|\Psi_{extreme} \rangle =\frac{1}{\sqrt{2}}\left(|N,0\rangle+|0,N\rangle\right)
\label{eqn:twowellcat1}
\end{equation}
where all the particles are simultaneously in the left and right
wells. Here the cat states are positional NOON states where $N$
particles occupy spatially separated modes in two wells. There can
also be NOON states of other kinds such as with $N$ particles
occupying the two quasi-momentum modes of counter-propagating
superfluid flow states in a rotating ring lattice that have been
proposed in Ref.~\cite{dunningham2006,nunnenkamp08}.

Anderson~\cite{manybody1964} showed that the Hamiltonian for a
system of two quantum fluids connected by a tunneling junction can
be described as a physical pendulum. For BEC in a lattice, in the
semiclassical limit valid for large $N$, the operators
$\hat{a_{i}}$ in Eq.~\ref{eqn:BHhamil} can be approximated by the
c-numbers $\sqrt{ n_{i}}e^{i\theta _{i}}$, where $n_{i}$ and
$\theta _{i}$ are the number and phase of particles in the $i$th
well. For a two site Bose-Hubbard model, the Hamiltonian then
turns into a classical Hamiltonian of a nonrigid physical pendulum
with the number and phase differences
($n=(n_L-n_R)/2$,$\theta=\theta_L-\theta_R$) between the wells as
conjugate variables.  The dynamics of double well BEC is then
described by a classical pendulum model that has been studied in
detail in Ref.~\cite{smerzi1997} and experimentally verified in
~\cite{albiez05}. Fig.~\ref{fig:phasespace}(a) shows the classical
energy contour showing the classical phase space structure. This
system has two fixed points -- (0,0) and (0,$\pi$).  The (0,0) is
a stable equilibrium, while the (0,$\pi$) is stable in the
$\pi$-state regime ($UN/J<1$) and unstable otherwise ($UN/J>1$).
Two types of pendulum motions -- oscillations and rotor motions
appear in the phase space, below and above the separatrix.
Question can be raised on how much of the classical phase space is
actually contained in a full quantum analysis. This was answered
in Ref.~\cite{mahmud2005} showing quantum-classical correspondence
between the classical pendulum and double well BEC.

Husimi probability distribution~\cite{mahmud2005,husimi40} can be
used to project, in a squeezed coherent state representation, the
classical phase space properties from quantum wavefunctions. In
$(n,\theta)$ representation, Husimi function is defined as
\begin{equation}
P_{j}(n,\theta)=\vert\langle\theta+in|\Psi_{j}\rangle\vert^2
\label{eqn:husimi1}
\end{equation}
where
\begin{equation}
\langle\theta+in|\Psi_{j}\rangle=\frac{1}{(\pi\kappa)^{1/4}}
\sum_{n^{\prime}=-N/2}^{N/2}c^{j}_{n^{\prime}}\exp\lbrack i\theta
n^{\prime}- \frac{(n^{\prime}-n)^2}{2\kappa}\rbrack
\label{eqn:husimi2}
\end{equation}
Here $n^{\prime}=\frac{n_L-n_R}{2}$, rather than being the simpler
left particle counter, and $c_{n^{\prime}}$ is the corresponding
Fock-state coefficient.  The `coarse-graining' parameter $\kappa$
determines the relative resolution in phase space in the conjugate
variables number and phase.

\begin{figure}[ht]
\includegraphics[width=0.45\textwidth,angle=0]{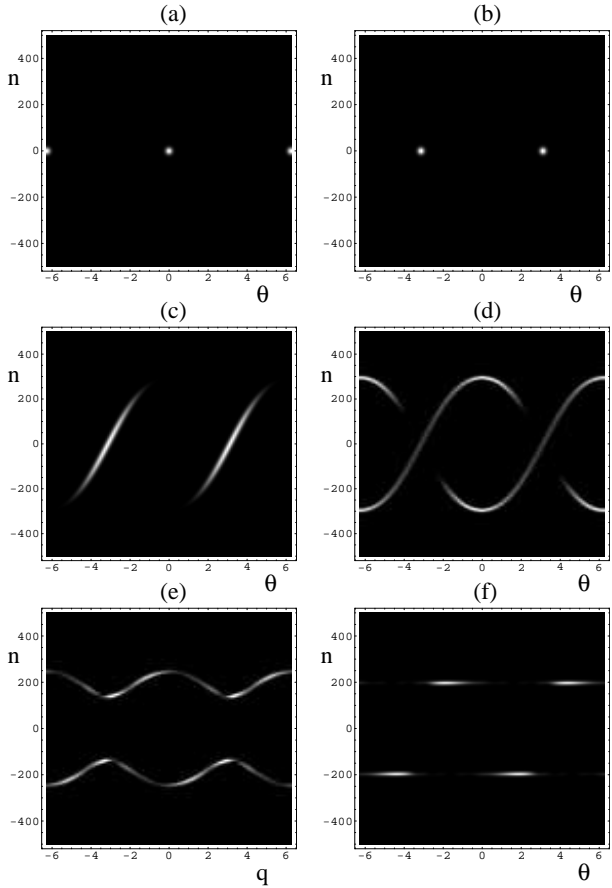}
\caption{\label{fig:husimi} A visual depiction of the underlying
physical principles for the generation of entangled states in a
double well. Quantum phase space pictures of Husimi projections
are plotted here - (a) the ground state at t=0, (b) $\pi$-phase
imprinted ground state at the hyperbolic fixed point, (c) at
t=0.012 ms the wavepacket is bifurcating along the separatrix, (d)
at t=0.019 ms it continues to move along the separatrix, (e) with
simultaneous ramping of the barrier the wavepacket splits
completely at t=0.49 ms, and (f) at t=2.84 ms a sharply peaked
entangled state is obtained which is a macroscopic superposition
of particles simultaneously in both wells.}
\end{figure}
\begin{figure}[ht]
\includegraphics[width=0.45\textwidth,angle=0]{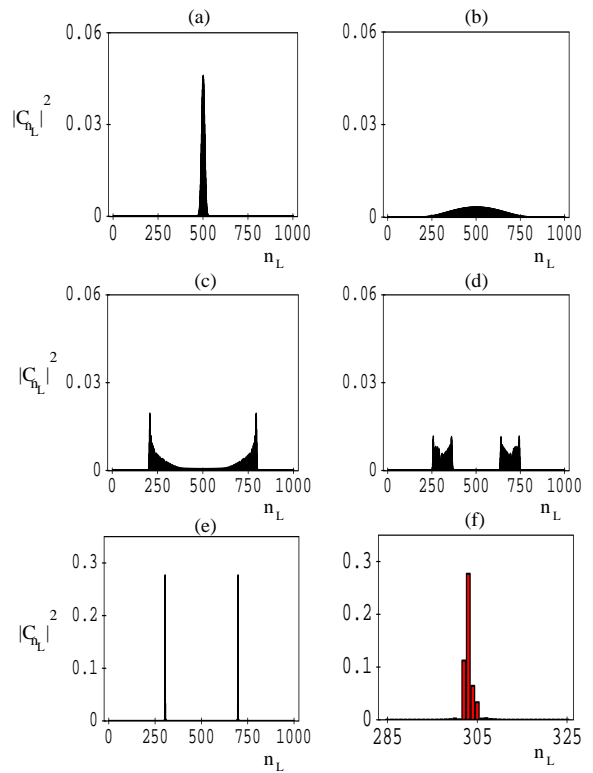}
\caption{\label{fig:twowellcat} Shown is the evolution to a cat
state in Fock space at the same time instants as in the previous
figure: (a) the phase imprinted ground state at t=0, (b) at
t=0.012 ms the Gaussian distribution broadens, (c) at t=0.019 ms
it bifurcates, (d) at t=0.49 ms it splits completely, (e) at
t=2.84 ms a highly entangled state is formed; (f) is a magnified
version of (e) showing the nonvanishing Fock state coefficients.}
\end{figure}

Figs.~\ref{fig:phasespace}(b)-(e) show representative Husimi
projections for 40 particles for the ground state, 6th , 12th and
35th states respectively. The ground state is a wave packet
centered on (0,0) in the classical phase space, with a finite
width in number and phase differences. The higher lying states in
(c) is a harmonic oscillator like state, (d) is a state which lies
on the separatrix and (e) is a cat state, which in classical
sense, is a superposition of clockwise and counter-clockwise
pendulum rotor states. After understanding this quantum-classical
correspondence, we can argue that a ground state wave packet
displaced in phase by $\pi$ and put onto the unstable fixed point
(0,$\pi$) would bifurcate along the separatrix and create a
superposition of two pendulum rotor states, if allowed to time
evolve. The $\pi$ displacement of the ground state can be
accomplished by phase imprinting one of the wells by an amount
$\pi$ which in experiments could be done by a phase engineering
technique that has been demonstrated in experiments investigating
solitons in the BEC~\cite{denschlag2000}.

For a concrete example of our method, we show results of a
numerical simulation in Figs.~\ref{fig:husimi} and
\ref{fig:twowellcat}.  Fig.~\ref{fig:husimi} demonstrates the
underlying physical principles quite visually in quantum phase
space using Husimi projections.  First in panel (a) we have a
ground state that is at the center of phase space, then in (b) we
displace this by $\pi$ in the horizontal direction so that it is
on an unstable equilibrium. Since it is no longer an eigenstate,
it will time evolve spontaneously, and in this case the
trajectories in phase space follow along the separatrix and
symmetrically spits in two directions as in (c) at t=0.012 ms.  In
this process the phase space points reach the top and bottom of
the separatrix as in (d) at t=0.019 ms. If we simultaneously
increase the barrier height during this time evolution, we can
completely split the top and bottom as shown in (e) at t=0.49 ms,
finally giving rise to a desired cat state in (f) at t=2.84 ms.
The times for this example of double well as well as three wells
in next section are given for a $^{87}$Rb condensate,
$\lambda=840$ nm, $a_{sc}=5.8$ nm, $J=0.04E_{R}e^{-\alpha}$ where
$E_{R}=\frac{\hbar^{2}k^{2}}{2m}$ is the recoil energy from
absorption of a photon, and $\alpha$ is a dimensionless parameter
that can be mapped onto the inverse tunneling rate, and taking
$U=0.04E_{R}$ as approximately constant for calculation purposes.
Here $\alpha$ varies as $\alpha=3+2t$. The evolution to an
entangled state is shown in Fock space coefficients in
Fig.~\ref{fig:twowellcat} -- the process of the formation of a
superposition state can also be understood from this. For a
two-component spinor Bose gas, Ref.~\cite{micheli03} shows the
generation of entangled state, where there is no need for initial
phase imprint. We demonstrate here the generation of number
entangled cat states of the NOON type; we do not discuss
generation of phase cat states~\cite{piazza08} that are
superpositions of many coherent phase states. Much of the
intuition for later sections is derived from an in depth study of
the double well as briefly described here and presented in detail
in~\cite{mahmud2005}.

\section{Entangled number states in three and four wells with small particle number}

\begin{figure}[ht]
\includegraphics[width=0.45\textwidth,angle=0]{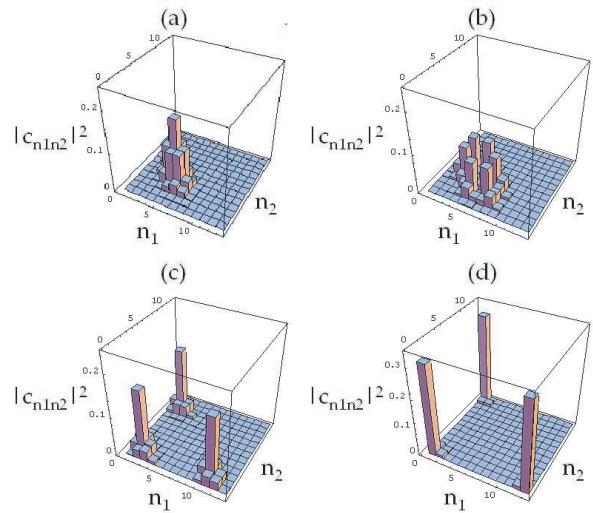}
\caption{\label{fig:stationary} (color online). Fock state
coefficients for 12 particles in three wells: (a) the ground
state, (b) 10th, (c) 76th and (d) 91st, the highest state. The
ground state has a Gaussian shape, while higher lying states are
entangled number states. $n_1$ and $n_2$ are the Fock state
indices and the vertical axis shows probabilities. Points beyond
the cross-diagonal are unphysical.}
\end{figure}

In order to gain insight into the multiwell Bose-Hubbard model, we
first analyze the quantum mechanical properties of the simplest
multiwell potential, $M=3$, assuming three symmetric wells in a
circular array~\cite{circularandlineararray}.

The state vector for three wells is a superposition of all the
number states
\begin{equation}
|\Psi _{i}\rangle
=\sum_{n_{1},n_{2}=0}^{N}c_{n_{1},n_{2}}^{(i)}(t)|n_{1},n_{2},n_{3}\rangle
\label{eqn:basisthree}
\end{equation}
Here $n_{1}$, $n_{2}$, and $n_{3}={N}-n_{1}-n_{2}$ are the number
of particles in each of the three wells. Fig.~\ref{fig:stationary}
shows the Fock space probabilities,
$\left|c_{n_{1},n_{2}}^{(i)}\right| ^{2}$, for representative
stationary states for $N=12$ and $\alpha =0$ ($U/J=1$). $n_1$ and
$n_2$ are the Fock state indices and the vertical axis shows
probabilities. The ground state in Fig.~\ref{fig:stationary}(a) is
a broad Gaussian while the higher lying states,
Figs.~\ref{fig:stationary}(c)-(d), are number entangled states of
increasing extremity corresponding to increasing numbers of
particles simultaneously in all three wells, the highest of which
in panel (d) approximates an extreme superposition state of the
form $|N,0,0\rangle+|0,N,0\rangle+|0,0,N\rangle$. Note that there
are still some nonvanishing Fock space components. As $U/J$
increases the highest lying state approaches the extreme state of
Eq~\ref{eqn:extremeCat}. The number of non vanishing Fock state
coefficients determines sharpness, and thus (d) is sharper than
(c).

\begin{figure}[ht]
\includegraphics[width=0.45\textwidth,angle=0]{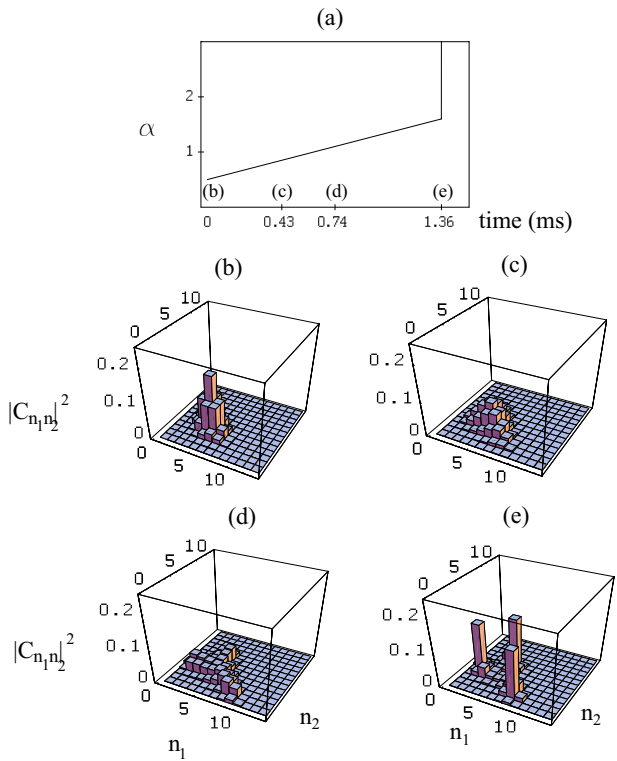}
\caption{\label{fig:evolving} (color online). Evolution to an
entangled Fock space state for three wells: (a) barrier ramp
showing the location of the following time evolved states: (b)
initial state, (c) at 0.43 ms the Gaussian distribution broadens,
(d) at 0.74 ms the distribution is `splitting', (e) A three-peaked
state is formed at 1.36 ms; a macroscopic superposition of
definite number of particles simultaneously in all three wells. A
comparison with the double well Fock state evolution in
Fig.~\ref{fig:twowellcat} shows that the physical mechanism for
generating entangled states is the same in a double well and
multiple wells.}
\end{figure}
\begin{figure}[ht]
\includegraphics[width=0.45\textwidth,angle=0]{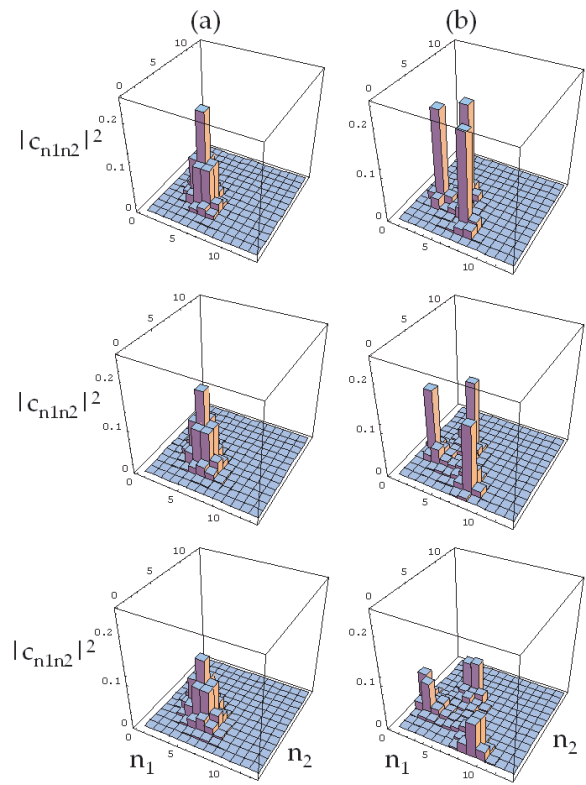}
\caption{\label{fig:tunable} (color online). Entangled states
evolved from ground states with different initial squeezing. Row
(1) shows the states with $\alpha=1.0+t$: (a) initial ground state
and (b) final state. Row (2) is for $\alpha=0.5+t$ and (3) is for
$\alpha=t$. Column (b) gives the states at t=1.85 ms, t=1.36 ms,
and t=0.99 ms respectively. The initial squeezing of the ground
state thus determines the extremity of the resulting entangled
states.}
\end{figure}

It is unlikely that such maximally entangled states can be
generated via a sequence of single particle excitations. They may
however, be dynamically generated via phase engineering from the
appropriate ground state, as elucidated in previous section for
the double well. Writing phases on part of a condensate is
experimentally feasible via interaction with a far off-resonance
laser~\cite{denschlag2000}, and is assumed to be sudden with
respect to the dynamics of the condensate. Mathematically, this
corresponds to multiplying the coefficients in an expansion of the
type of Eq.~\ref{eqn:fockgen} by $e^{in_{i}\theta _{i}}$, where
$|n_{1},n_{2},...n_{i},...\rangle$ is the corresponding Fock
state, and $\theta _{i}$ is the phase for particles in the $i$th
well.

Entangled state generation, obtained via integration of the linear
time-dependent Schr\"{o}dinger equation, is shown in
Fig.~\ref{fig:evolving}, following phase imprinting of an initial
phase difference of $\frac{2\pi}{3}$ between the neighboring
wells, and a simultaneous linear ramping of the barrier as
$\alpha= 0.5+t$, as shown in Fig.~\ref{fig:evolving}(a) (t here is
dimensionless). Panels \ref{fig:evolving}(b) shows the initial
ground state; \ref{fig:evolving}(c) at time 0.43 ms, the
distribution broadens; \ref{fig:evolving}(d) at 0.74 ms, in the
process of splitting the state towards the three corners; and
\ref{fig:evolving}(e) at 1.36 ms a sharp, although not extreme,
entangled number state with its signature of three major non
vanishing expansion coefficients.

When an appropriately entangled state is reached the barrier is
suddenly raised to halt further evolution in n-space. For the
parameter values used here, a simple time evolution without any
change of barrier also produces an entangled state, however
barrier ramping is used here to sharpen the resulting state, and
completely split Fock state coefficients into three parts. Control
of the extremity of the states can be achieved by choice of the
initial barrier height controlling the initial squeezing of the
ground state. This is demonstrated in Fig.~\ref{fig:tunable},
where different initial squeezing have been used for rows (1), (2)
and (3). The columns show: (a) the ground state, and (b) the final
state at the end of the barrier ramping. It is important to be
able to tune to less extreme entangled states, as such states are
more robust to loss and
decoherence~\cite{mahmud2005,mahmudThesis}. In our study here, we
show the proof of principles that cat states can be produced and
controlled. To be able to get extreme superpositions, analysis can
be made with optimal control theory on the correct parameters and
ramping to be used. Phase imprinting with a phase difference of
$\frac{4\pi}{3}$ produces an equivalent state, with different
phase space dynamics.

What is so special about phase imprinting $\frac{2\pi}{3}$?
Similar to the simple double well, the triple well, $M=3$, can be
thought of as two coupled pendulums~\cite{franzosi2003} with
complicated dynamics -- quantum and semiclassical aspects of three
well BEC have been elucidated in
Ref.~\cite{mossman06,trimborn09,viscondi10}. Here the
semiclassical conjugate variables are $\Delta n_{12}=n_1-n_2$,
$\Delta n_{23}=n_2-n_3$, $\Delta \theta_{12}=\theta_1-\theta_2$
and $\Delta \theta_{23}=\theta_2-\theta_3$. The unstable fixed
points in these conjugate variables are (0,0,$\frac{2k\pi
}{3}$,$\frac{2k\pi }{3}$), $k=1,2$. So, a phase imprint of
$\frac{2\pi}{3}$ or $\frac{4\pi}{3}$ puts the ground state
wavepacket on the unstable equilibrium, and subsequent dynamics
gives rise to these states.

\begin{figure}[ht]
\includegraphics[width=0.45\textwidth,angle=0]{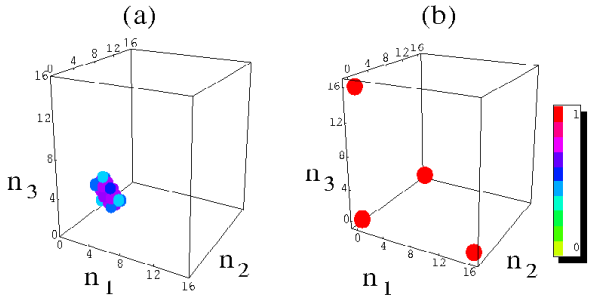}\caption{\label{fig:fourwell}
(color online). Four well stationary and time evolved states: (a)
ground state, (b) an entangled state evolved from the ground state
following a $\pi$ relative phase shift. The three dimensions show
the Fock state indices $n_1$, $n_2$ and $n_3$, probabilities are
shown in the color intensity scale. For graphical clarity, only
the points higher than $40\%$ of the highest probability are
shown, with the highest probabilities normalized to 1.}
\end{figure}
\begin{figure}[ht]
\includegraphics[width=0.45\textwidth,angle=0]{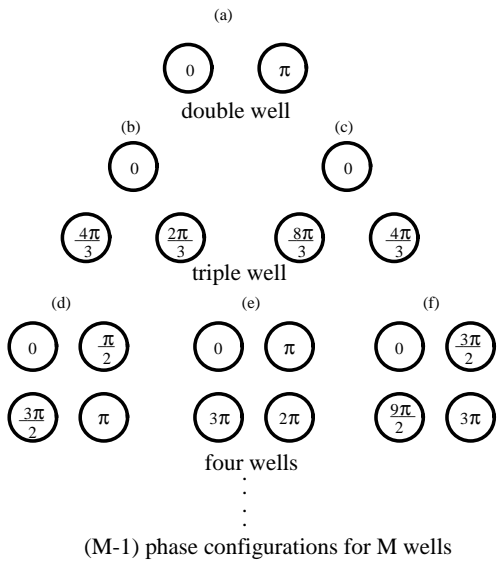}
\caption{\label{fig:configure} Initial phase configurations that
generate number entangled states. These phase differences are
motivated by the fixed point of the underlying semiclassical phase
space -- bifurcation characteristics of the unstable equilibrium
generating the macroscopic superpositions. Shown here are examples
for a double well (a), triple well (b and c), and four wells (d, e
and f). There are $M-1$ different possibilities for $M$ wells. The
known case of $\pi$ phase differences for an even number of wells
is seen to be a special case.}
\end{figure}

One important aspect of our method is the controllability of the
final state which works for multiple wells as illustrated in
Fig.~\ref{fig:tunable}. It works in three ways -- I) A
simultaneous ramping of the barrier with the natural dynamics at
the unstable fixed point has been empirically found to be useful
in directing the desired evolution of the wavepacket. II) Initial
barrier height, that is the initial squeezing, helps shape the
initial wave packet stretching it into different regions of
accessible phase space; and, III) the initial barrier height sets
the (negative) curvature of the potential at the hyperbolic fixed
point, controlling the rate of splitting of the wave packet.

We found that all the features of the double well entanglement
generation apply to the three well case, and thus, many of the
insights from the two and three well dynamics can be extended to
arbitrary number of wells in a circular array. Next, we explore it
for four wells. For four wells, the values of phase imprints that
take the ground state to fixed points are $\frac{\pi}{2}$, $\pi$
and $\frac{3\pi}{2}$. Fig.~\ref{fig:fourwell}(a) shows a typical
ground state in four wells that is an approximate Gaussian. After
a phase imprint of $\pi$ between neighboring wells, the ground
state can be evolved into an extreme entangled number state as
shown in Fig.~\ref{fig:fourwell}(b) for $N=16$, $U=0.01$ $E_R$,
$J=0.04$ $E_R$ $e^{-\alpha}$, with $\alpha=0.175$, and assuming
the case of $^{87}$Rb in the previous example. Phase difference of
$\pi$ between neighboring wells is equivalent to writing
alternating $\pi$ phases on the lattice. The other fixed point
dynamics of $\frac{\pi}{2}$ and $\frac{3 \pi}{2}$ also lead to
symmetric states but do not lead to the highly entangled states of
the kind we discuss in this article.

In Ref.~\cite{polkovnikov2002}, a truncated Wigner approximation
was used to study an alternating $\pi$ phase difference dynamics
for even number of wells. In comparing our four well results to
theirs, we have done an exact time evolution study, and find that
the $\pi$ configuration in an even number of wells that they have
identified is just a special case of many phase imprint dynamics
that could generate interesting correlated states in multiple
wells. Their changes in system parameters is to drive the system
from stability to a regime of instability. On the other hand, we
take our system to be in the unstable regime and demonstrate the
controllability of entangled states with barrier manipulation;
potentially useful for experimental detection.

For the two, three, and four wells, we find $M-1$ distinct phase
differences between the neighboring wells for the multi-well fixed
points~\cite{franzosi2003}. These are given by a general formula
$\frac{2\pi j}{M}$ where $j=1,2,....M-1$, with $M$ being the
number of wells, which gives a $\pi$ phase difference for the
$M=2$ double well, a $\frac{2\pi}{3}$ and $\frac{4\pi}{3}$ phase
difference for the $M=3$ triple well, and a $\frac{\pi}{2}$,
$\pi$, and $\frac{3\pi}{2}$ phase difference for the $M=4$
quadruple well configuration -- we investigated the dynamics
generated by all of these phase difference imprints. Note that the
total change in phase in the circular loop is a multiple of
$2\pi$, a vortex like condition. We thus propose a general formula
for $M$ wells,
\begin{equation}
\Delta \theta =\frac{2\pi j}{M}\text{,}
\label{eqn:phase}
\end{equation}
for the constant phase offset between neighboring wells leading to
the dynamical generation of entangled states. Here $j=1,2,..,M-1$,
and Eq.~(\ref{eqn:phase}), being valid for any number of wells,
even or odd, provides a substantial generalization of the $\pi$
phase offset mentioned in Ref.~\cite{polkovnikov2002}, which is
valid only for the special cases of an even number of wells and
for $j=M/2$. Fig.~\ref{fig:configure} shows the phase
configurations of Eq.~(\ref{eqn:phase}). The multiplicity of
Eq.~(\ref{eqn:phase}) is prominent for large number of wells, e.g.
for 12 wells, there are 11 phase offset possibilities. Symmetries
may prevent all the imprinting offsets of Eq.~\ref{eqn:phase} from
generating independent dynamics.

Although detailed work on four and eight wells was done more
recently and is being presented here as a coherent whole for
multiple wells, the work on three well was done much earlier as
documented in Ref.~\cite{mahmudThesis,mahmud04}. Many studies of
quantum, semiclassical and entanglement aspects of three coupled
BECs have appeared since then, that continue to explore the rich
yet simpler lattice physics of the system of three well BEC.

\section{Large entangled number states in multiple wells}

In this section we investigate the creation of large entangled
number states where there are a large number of particles in each
well and/or a large number of wells. Following the techniques
described above and in~\cite{leung2007}, we explored the creation
of entangled number states in four and eight wells with periodic
boundary conditions. We present results from our simulations of
entangled number states in four wells in the Section V A, eight
wells in V B, and in V C we analyze the types of entangled states
generated and describe a new type of entangled number state.

In systems with wells $M > 4$, visualization of multi-dimensional
entangled states becomes more difficult. To facilitate the
visualization we introduce the joint probability function
\begin{equation}
P(n_\beta,n_\gamma,t)=\frac{1}{M}\sum_{n_i=0, \forall i \neq
\beta,\gamma}^{N}|c_{n_1,n_2,\ldots,n_i,\ldots,n_M}(t)|^2
\label{eqn:jointprob}
\end{equation}
where the sum over $n_i$ does not include $n_\beta$ or $n_\gamma$,
as they are held fixed and particle conservation requires a fixed
total number of particles.  $P(n_\beta,n_\gamma,t)$ shows the
probability of finding $n_\beta$ particles in $\beta$th well
simultaneously with finding $n_{\gamma}$ particles in the
$\gamma$th well.

\subsection{Large entangled number states in four wells}

%
\begin{figure}[ht]
\begin{center}
\includegraphics[width=0.45\textwidth,angle=0]{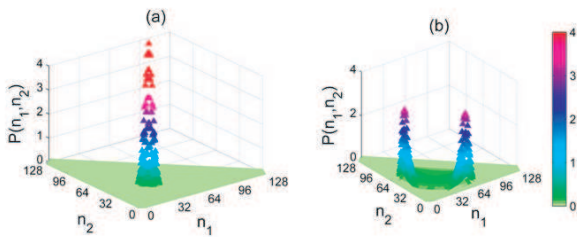}
\end{center}
\caption{\label{fig:128in4} (color online). Ground and entangled
number state for 128 particles in four wells: (a) Ground state and
(b) entangled number state at $t=2.40$ ms for 128 particles in
four wells, $U=0.01 E_R$, $J=0.04e^{-\alpha}E_R$ with
$\alpha=0.175$. The time shown in (b) represents the earliest time
when maximum probability occurs for an entangled number state.
Shown is the joint probability function $P(n_1,n_2)$ versus the
number of particles in wells 1 and 2 on the $x$ and $y$ axes,
respectively.}
\end{figure}
\begin{figure}[ht]
\begin{center}
\includegraphics[width=0.45\textwidth,angle=0]{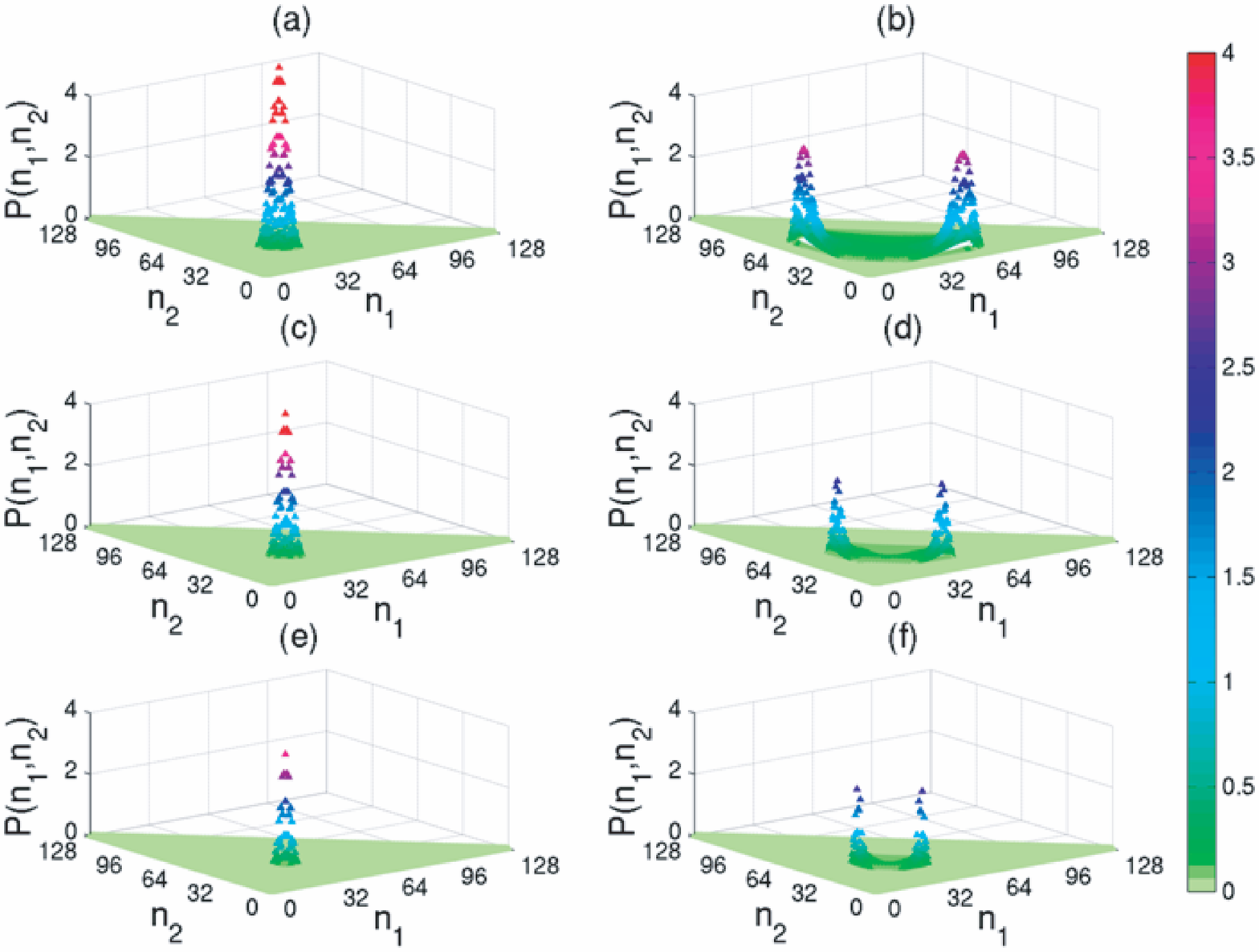}
\end{center}
\caption{\label{fig:varyJ} (color online). Varying the tunneling
parameter, $J$, for four wells: (a) Ground state and (b) entangled
number state for 128 particles in four wells with $U=0.01 E_R$ and
$J=0.04e^{-\alpha} E_R$ with $\alpha=0.175$ and $t=2.40$ ms,
$\alpha=1.175$ and $t=3.65$ ms, and $\alpha=2.175$ and $t=5.70$ ms
in rows 1, 2, and 3. This shows that initial squeezing can be used
to control the final entanglement as have been shown for three
wells in Fig.~\ref{fig:tunable} and double wells in
Ref.~\cite{mahmud2005}}
\end{figure}

The results of our investigation of the creation of entangled
number states in four wells with a large number of particles in
each well are provided below.  The regime we explore in four wells
is similar to the Kasevich~\cite{orzel2001} experiments and we
report results using experimental parameters relevant to their
work for a condensate made of $^{87}Rb$ with $\lambda=840$ nm and
$a_{sc}=5.8$ nm.  We assume an intial phase offset of $\pi$
between the wells. Fig.~\ref{fig:128in4} shows the joint
probability function for (a) the ground state at $t=0$ and (b)
entangled number state at time $t=2.40$ ms for 128 particles in
four wells.  In this figure and all following the time depicted
represents the earliest time when maximum probability for an
entangled state occurs.  The ground state has a Gaussian
distribution centered around the state where the particles are
equally distributed between the wells.  The joint probability
function for the entangled number state has two peaks one where
there are 57 particles in the $\beta=1$ well and 7 particles in
the $\gamma=2$ well and another peak of equal probability where
there are 7 particles in the $\beta=1$ well and 57 particles in
the $\gamma=2$ well.  Besides this highest probability state,
there are other non-vanishing coefficients that are smaller and
distributed around this peak.

Fig.~\ref{fig:varyJ} shows the effect of varying $J$, the
tunneling parameter. Column (a) shows the joint probability
function for the ground state at $t=0$ and column (b) shows an
entangled number state at $t=2.40$, $3.65$, and $5.70$ ms in rows
1, 2, and 3, respectively.  The tunneling parameter is set to
$J=0.04e^{-\alpha}E_R$ with $\alpha=0.175$, $\alpha=1.175$, and
$\alpha=2.175$ for rows 1, 2, and 3 respectively.  Increasing
$\alpha$ corresponds to decreasing tunnelling and correlates with
increased time to evolve into the entangled number state.
Increased $\alpha$ also means starting with a state with larger
number squeezing, and thus we see in panels (a), (b) and (c) that
the extremity of the final cat state is determined by initial
squeezing. This same behavior was shown with double
well~\cite{mahmud2005}, and with three wells in the previous
section, and thus we show that it is a general characteristic of
multi-well cat states. We should emphasize the finding that to
obtain a more extreme cat state, the initial barriers have to be
low.

\begin{figure}[ht]
\begin{center}
\includegraphics[width=0.45\textwidth,angle=0]{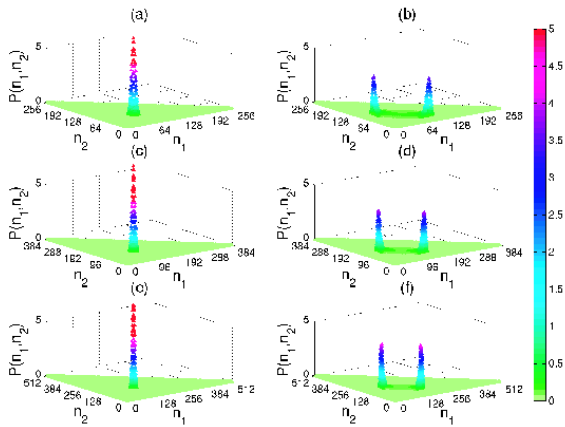}
\end{center}
\caption{\label{fig:varyN} (color online). Varying number of
particles for four wells: (a) Ground state and (b) entangled
number state with $U=0.01E_R$ and $J=0.04e^{-\alpha}E_R$ with
$\alpha=0.175$ for 256 particles at $t=1.74$ ms, 384 particles at
$t=1.47$ ms, and 512 particles at $t=1.30$ ms in four wells for
rows 1, 2, and 3. For the same barrier height, higher number of
particles give rise to less extreme superpositions.}
\end{figure}

Fig.~\ref{fig:varyN} illustrates the effect of varying the number
of particles. Column (a) shows the ground state at $t=0$ and
column (b) shows the entangled number state for  256 particles at
$t=1.74$ ms, 384 particles at $t=1.47$ ms, and 512 particles at
$t=1.30$ ms.  Increasing the particle number raises the strength
of the interaction term in the BH Hamiltonian and results in
decreased tunneling, explaining the increased squeezing of the
ground and entangled number states with increased particle number.

Our simulations indicate that entangled number states evolve
through natural time evolution of the BEC with an initial phase
offset of $\pi$ between the wells even with a large number of
particles in each well.  The extremity of the entangled number
state can be controlled by varying the tunneling parameter or the
number of particles.

\subsection{Entangled number states in eight wells}

%
\begin{figure}[ht]
\begin{center}
\includegraphics[width=0.45\textwidth,angle=0]{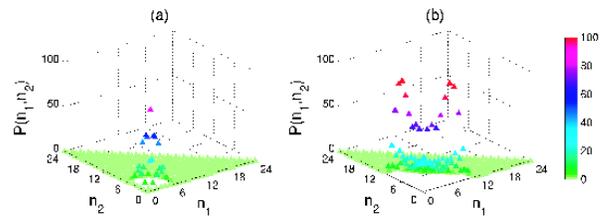}
\end{center}
\caption{\label{fig:eightwellcat} (color online). 24 particles in
eight wells: (a) Ground state and (b) entangled number state at
$t=2.79$ ms for 24 particles in eight wells with $U=0.01 E_R$ and
$J=0.04 e^{-\alpha} E_R$ with $\alpha=0.175$.}
\end{figure}
We present the dynamic creation of entangled number states in
eight wells with just a few particles per well; this regime is
similar to the regime studied by Greiner~\cite{greiner2002} and we
report our results using experimental parameters relevant to their
work; $^{87}Rb$, $\lambda=852$ nm, $a_{sc}=5.8$ nm.
Fig.~\ref{fig:eightwellcat} shows (a) ground state and (b) the
time evolved entangled number state for 24 particles in eight
wells with $J=0.04 e^{-\alpha}E_R$ with $\alpha=0.175$ and
$U=0.01E_R$ with an initial phase offset of $\pi$ between the
wells. Our investigations show that entangled number states can
also be realized in a larger number of wells with a small number
of particles in each well.

\subsection{Analysis of entangled number states: a new type of entangled state}

The extreme entangled state for a multiwell BEC takes the form of
Eq.~\ref{eqn:extremeCat}. And the less extreme entangled states in
multiple wells can have the approximate form
\begin{eqnarray}
|\Psi \rangle_{\{N, M \}^{non-extreme}} &=&\frac{1}{\sqrt{M}}(|N-jn,n,\ldots,n\rangle \nonumber \\
&&+|n,N-jn,n,\ldots,n\rangle \nonumber \\
&&+\ldots+|n,\ldots,N-jn\rangle) \label{eqn:squeezedCat}
\end{eqnarray}
where $n<<N$ and $j=M-1$. The four well cat state shown in
Fig.~\ref{fig:fourwell}(b) is an extreme cat state in the
approximate form $|\Psi
\rangle=\frac{1}{\sqrt{4}}\left(|16,0,0,0\rangle+|0,16,0,0\rangle+|0,0,16,0\rangle+|0,0,0,16\rangle
\right)$. An example of a less extreme cat state would be of the
form $|\Psi
\rangle=\frac{1}{\sqrt{4}}\left(|13,1,1,1\rangle+|1,13,1,1\rangle+|1,1,13,1\rangle+|1,1,1,13\rangle
\right)$. Thus, one would expect a similar form for the large
entangled states described in this section, however this is not
what we found, as is shown below.

The states in Fig.~\ref{fig:128in4} with the highest probability
at $t=0$ and $t=2.40$ ms are shown in the first two entries of
Table~\ref{tab:varyJ} and indicate there are 32 particles in each
of the wells in the ground state and 57 particles simultaneously
in all four wells in the entangled number state. The highest
probability states in Fig.~\ref{fig:varyJ} with varying $J$ are
shown in Table~\ref{tab:varyJ}. Table~\ref{tab:varyJ} shows the
ground states with the particles equally distributed between the
wells and the entangled states with 57, 48, and 42 particles in
all the wells. The highest probability states for
Fig.~\ref{fig:varyN} with varying $N$ are shown in
Table~\ref{tab:varyN}, and Table~\ref{tab:24in8} shows the highest
probability states for the ground and entangled number states for
the eight well system shown in Fig.~\ref{fig:eightwellcat}. We
would like to emphasize that the listed state is the highest
coefficient state, and there are other lower coefficient states
around this.

\begin{figure}[ht]
\begin{center}
\includegraphics[width=0.45\textwidth,angle=0]{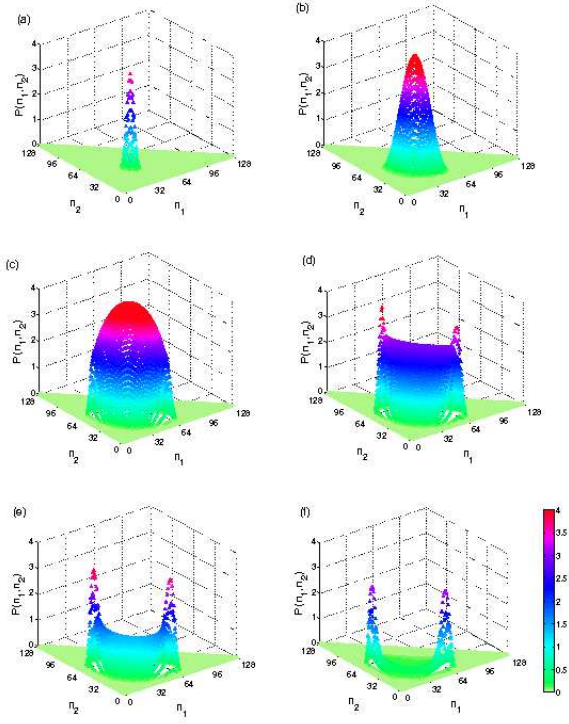}
\end{center}
\caption{\label{fig:fourEvol} (color online). Evolution of
entangled state for 128 particles in four wells, shown in the
joint probability distribution space of Eq.~\ref{eqn:jointprob}
for sites 1 and 2. Shown are (a) at t=0, the ground state, (b) at
t=1.22 ms, the ground state broadens, (c) at t=1.71 ms, (d) at
t=1.84 ms the state is beginning to split just as it does in a
double well and triple well, (e) at t=1.96 ms, and (f) at t=2.42
ms, the final entangled state is formed. The evolution here
further illustrates that the underlying physical mechanism for the
creation of cat states is the same in a double well and multiple
wells.}
\end{figure}
%

\begin{table}[h]
\begin{center}
\caption[Highest probability states for 128 particles in four
wells with varying $J$]{Highest probability states for 128
particles in four wells with $U=0.01 E_R$ and varying $J$, where
$J=0.04 e^{-\alpha} E_R$. These states correspond to
Fig.~\ref{fig:varyJ}. It should be emphasized that all the Tables
in this article show only the highest state; however, there are
other non-vanishing coefficients as well as evident from
Fig.~\ref{fig:varyJ}. Those coefficients are smaller and spread
out around the highest probability peak.}
\begin{tabular}{rrl}
\hline
\hline
t (ms)&$\alpha$&highest probability state \\
\hline
0.0&0.175&$|32,32,32,32\rangle$\\
2.4&0.175&$\frac{1}{\sqrt{2}}(|57,7,57,7\rangle+|7,57,7,57\rangle)$\\
0.0&1.175&$|32,32,32,32\rangle$\\
3.65&1.175&$\frac{1}{\sqrt{2}}(|48,16,48,16\rangle+|16,48,16,48\rangle)$\\
0.0&2.175&$|32,32,32,32\rangle$\\
5.7&2.175&$\frac{1}{\sqrt{2}}(|42,22,42,22\rangle+|22,42,22,42\rangle)$\\
\hline
\hline
\\
\end{tabular}
\label{tab:varyJ}
\end{center}
\end{table}

\begin{table}[h]
\begin{center}
\caption[Highest probability states in four wells with varying
$N$]{Highest probability states in four wells with $J=0.04
e^{-0.175}E_R$, $U=0.01 E_R$ and varying $N$. These states
correspond to Fig.~\ref{fig:varyN}.}
\begin{tabular}{rrl}
\hline
\hline
t (ms)&$N$&highest probability state\\
\hline
0.0&256&$|64,64,64,64\rangle$\\
1.74&256&$\frac{1}{\sqrt{2}}(|102,26,102,26\rangle+|26,102,26,102\rangle)$\\
0.0&384&$|96,96,96,96\rangle$\\
1.47&384&$\frac{1}{\sqrt{2}}(|143,49,143,49\rangle+|143,49,143,49\rangle)$\\
0.0&512&$|128,128,128,128\rangle$\\
1.30&512&$\frac{1}{\sqrt{2}}(|183,73,183,73\rangle+|73,183,73,183\rangle)$\\
\hline
\hline
\\
\end{tabular}
\label{tab:varyN}
\end{center}
\end{table}

\begin{table}[h]
\begin{center}
\caption[Highest probability states for in eight wells]{Highest
probability states in eight wells with $J=0.04 e^{-0.175}E_R$,
$U=0.01 E_R$. These states correspond to
Fig.~\ref{fig:eightwellcat}.}
\begin{tabular}{rl}
\hline
\hline
t (ms)&highest probability state\\
\hline
0.0&$|3,3,3,3,3,3,3,3\rangle$\\
2.79&$\frac{1}{\sqrt{2}}(|6,0,6,0,6,0,6,0\rangle+|0,6,0,6,0,6,0,6\rangle)$\\
\hline
\hline
\\
\end{tabular}
\label{tab:24in8}
\end{center}
\end{table}

Based on the empirical evidence of the entangled number states
shown in this section we find the entangled number states for
large $N/M$ and large $M$ tend towards the approximate form

\begin{eqnarray}
|\Psi \rangle &=&\left(\frac{1}{\sqrt{2}} \right)|\frac{2N}{M}-n,n,\frac{2N}{M}-n,\ldots,n\rangle \nonumber \\
&& + |n,\frac{2N}{M}-n,n,\ldots,\frac{2N}{M}-n\rangle
\label{eqn:multiwellcat2}
\end{eqnarray}

where $n < N$ and there are $\frac {2N}{M}-n$ particles
simultaneously in all $M$ wells. We say `approximate form' because
although the highest probability state has this form, there are
many non-vanishing coefficients as well with smaller amplitudes,
and spread out around the highest probability peak. The initial
parameters of the system and dynamics determines the relative
amplitudes of the non-vanishing coefficients for the final
entangled states generated in our model.

To our knowledge at the time of writing, number entangled states
in the form of Eq.~\ref{eqn:multiwellcat2} have not been discussed
elsewhere and represent a new type of entanglement. Well known
entangled states include Greene, Horne, Zeilnger
states~\cite{greenberger1990} shown in Eq.~\ref{eqn:ghz} which for
two wells and $N$ particles is an extreme entangled state of the
form, $(1/\sqrt{2})(|N,0\rangle+|0,N\rangle$), also known as NOON
states. Another type of state is the W-state~\cite{dur00}, which
for $N$ qubits has the form,
$(1/\sqrt{N})(|10..00\rangle+|01..00\rangle+..+|00..01\rangle)$.
Because of the `twin-like' nature, we refer to the number
entangled states of Eq.~\ref{eqn:multiwellcat2} as Siamese states.
For the case of qubits, W state was found to be more robust than
GHZ state. For multi-dimensional quDits, the Siamese states
introduced here may have different properties than other forms of
number entangled states. We have only shown here a way of
generating such novel states with cold atoms in a lattice without
yet knowing their properties.

Tables~\ref{tab:varyNsiamese} and \ref{tab:varyUsiamese}
illustrate the relationship between particle number and
interaction energy and these new types of entangled states.
Table~\ref{tab:varyNsiamese} shows that when we increase the
number of particles, the Fock state number distribution in the
final evolved state slowly turn into twin-like states. Same
observation can also be made from the increase of interaction
parameter $U$. Since particle number increment is effectively an
increase in effective interaction, these results complement each
other. Since in the Bose-Hubbard model the properties are
determined by the ratio of $U/t$, we can say that starting with a
small value of tunneling or a high barrier (highly squeezed state)
is a prerequisite for obtaining these states. This actually makes
the distinction very clear for the four well results, since we
have shown for double well and three well, that to get toward an
extreme cat state limit we need to start in the opposite limit of
low barrier (gaussian state).

Fig.~\ref{fig:fourEvol} shows evolution of a four well state, for
a twin-like state. In the graphical representation of two-site
joint probability function, a twin-like state and a four-legged
cat state cannot be differentiated. For that we look at the
coefficients of different Fock states to verify that this is a
twin-like state, with the highest probability state being
$(|57,7,57,7\rangle$+$|7,57,7,57\rangle)$, with other
non-vanishing coefficients that are smaller.
Fig.~\ref{fig:fourEvol} presents the general picture of how a four
well state evolves -- (a) shows the phase imprinted ground state,
(b) and (c) shows the broadening of the Gaussian distribution, (d)
and (e) shows the process of splitting into superpositions, and
(f) shows the final evolved state. This entangled number state
example is for 128 particles, and $U=0.01E_R$,
$J=0.04e^{-\alpha}E_R$ with $\alpha=0.175$.
Figure~\ref{fig:fourEvol} can be compared with
Figs.~\ref{fig:twowellcat} and \ref{fig:evolving} for double well
and triple well respectively. We see that the dynamics of
entangled state generation in a double well and multiple well
follow similar properties -- a gaussian ground state once phase
imprinted to an unstable equilibrium, evolves first by broadening
the gaussian, and then slowly splitting symmetrically to a
macroscopic superposition state.

The detailed mechanism whereby the entangled states of
Eq.~\ref{eqn:extremeCat} cross over to the new type of entangled
states, with alternating populations as observed here, is not
fully understood. Nor it it obvious why other types of
entanglement pairings and types of alternation do not seem to
occur. However, the most likely explanation of the cross-over
effect itself will simply involve the fact that the entangled
states of Eq.~\ref{eqn:extremeCat} have the maximal possible
energies for the system: the proposed types of phase engineering
explored here simply do not lead, as $N$ and $U$ increase, to
energies high enough to allow simple time evolution into entangled
states of such high energy. In terms of the underlying classical
phase space dynamics, we can say that the characteristics of the
unstable equilibrium (0,0,0,$\pi$,$\pi$,$\pi$) (in the conjugate
variables $n_{12}$, $n_{23}$, $n_{34}$, $\phi_{12}$, $\phi_{23}$,
$\phi_{34}$) changes as $U$ and $N$ is increased. For the double
well case we know that $UN/J>1$ makes the $\pi$-phase an unstable
equilibrium, and stable otherwise. Similarly, here the combination
of values of $U$ and $N$ changes the scenario and beyond a certain
value of $UN$, a complicated phase space dynamics constrains the
motion in phase space in such a way that only certain region is
accessible. Since we have presented quantum dynamics of multi-well
in this paper based on intuition from its semiclassical aspects
without a detailed study of dynamics in the underlying classical
phase space as was done for the double well~\cite{mahmud2005}, we
cannot quantify our findings.

\begin{table}[h]
\begin{center}
\caption[Entangled states as a function of $N$]{Highest
probability states in four wells with $J=0.04 e^{-\alpha} E_R$
with $\alpha=0.0$, $U=0.01 E_R$ and varying $N$.}
\begin{tabular}{rrrrl}
\hline
\hline
t(ms)&$N$&highest probability state\\
\hline
7.69&20&$\frac{1}{\sqrt{4}}(|20,0,0,0\rangle+|0,20,0,0\rangle+$\\
&&$|0,0,20,0\rangle+|0,0,0,20\rangle)$\\
6.39&24&$\frac{1}{\sqrt{4}}(|23,0,1,0\rangle+|0,23,0,1\rangle+$\\
&&$|1,0,23,0\rangle+|0,1,0,23\rangle)$\\
6.24&28&$\frac{1}{\sqrt{4}}(|26,0,2,0\rangle+|0,26,0,2\rangle+$\\
&&$|2,0,26,0\rangle+|0,2,0,26\rangle)$\\
5.38&32&$\frac{1}{\sqrt{4}}(|27,0,5,0\rangle+|0,27,0,5\rangle+$\\
&&$|5,0,27,0\rangle+|0,5,0,27\rangle)$\\
2.91&64&$\frac{1}{\sqrt{2}}(|32,0,32,0\rangle+|0,32,0,32\rangle)$\\
2.15&128&$\frac{1}{\sqrt{2}}(|59,5,59,5\rangle+|5,59,5,59\rangle)$\\
\hline
\hline
\\
\end{tabular}
\label{tab:varyNsiamese}
\end{center}
\end{table}

\begin{table}[h]
\begin{center}
\caption[Entangled states as a function of $U$]{Highest
probability states in four wells with $J=0.04 e^{-\alpha} E_R$
with $\alpha=0.0$, $N=32$ and varying $U$ with $U=u(0.01 E_R)$.}
\begin{tabular}{rrrrl}
\hline
\hline
t(ms)&$u$&highest probability state\\
\hline
5.70&0.15&$\frac{1}{\sqrt{4}}(|32,0,0,0\rangle+|0,32,0,0\rangle+$\\
&&$|0,0,32,0\rangle+|0,0,0,32\rangle)$\\
5.38&0.25&$\frac{1}{\sqrt{4}}(|27,0,5,0\rangle+|0,27,0,5\rangle+$\\
&&$|5,0,27,0\rangle+|0,5,0,27\rangle)$\\
2.91&0.50&$\frac{1}{\sqrt{4}}(|18,0,14,0\rangle+|0,18,0,14\rangle+$\\
&&$|14,0,18,0\rangle+|0,14,0,18\rangle)$\\
2.47&0.75&$\frac{1}{\sqrt{4}}(|17,1,13,1\rangle+|1,17,1,13\rangle+$\\
&&$|13,1,17,1\rangle+|1,13,1,17\rangle)$\\
2.08&1.00&$\frac{1}{\sqrt{4}}(|16,2,12,2\rangle+|2,16,2,12\rangle+$\\
&&$|12,2,16,2\rangle+|2,12,2,16\rangle)$\\
1.66&1.25&$\frac{1}{\sqrt{4}}(|14,2,14,2\rangle+|2,14,2,14\rangle)$\\
\hline
\hline
\\
\end{tabular}
\label{tab:varyUsiamese}
\end{center}
\end{table}

\section{Conclusions}

We have demonstrated phase engineering schemes for the generation
of entangled number states in multiple wells. These states
represent a superposition of particles in a large number of
spatial locations -- a multi-positional generalization of double
well NOON states. It is shown that entangled number states can be
evolved from phase imprinted ground states of BEC in multiple
wells, and the number of particles participating in the
entanglement can be controlled by varying the height of the
barrier between the wells, controlling the rate of ramping, and/or
the number of particles. The less extreme entangled number states,
which we show how to create in a controlled way, represent
macroscopic superposition states that are more robust to
experimental conditions and particle loss. We demonstrated our
scheme with small number of particles in a small lattice, large
number of particles per well, as well as with a large number of
wells. Our investigations for four wells revealed a new type of
number entangled state where a twin like state is formed, which we
refer to as Siamese states. We demonstrated a relationship between
the particle number and interaction strength and the formation of
these new types of entangled states.

The physical mechanism by which these states are generated can be
understood in terms of the underlying semiclassical phase space.
Ground state put on an unstable equilibrium splits the wavepacket
symmetrically to create these states. We briefly describe our
earlier work on double well~\cite{mahmud2005,mahmudThesis} to
demonstrate this in the semiclassical phase space, and to motivate
and generalize the study to larger number of wells and lattices.
The significance of the phase imprint values of $2\pi/3$ for three
wells, and $\pi$ for four wells is explained this way. For the
initial phase difference between the neighboring wells required to
create these states, we presented a novel series of formulae that
is valid for any number of wells, even or odd.

The creation, characterization, and applications of
multidimensional/multipositional Schr\"{o}dinger cat states of
atoms discussed in this article remain largely unexplored
experimentally, and the theoretical ramifications of such states,
should they be easily produced, are just emerging.

\begin{acknowledgments}
This work was supported by NSF grant PHY-0140091 and PHY-07-03278
and MAL gratefully acknowledges support from the DOE computational
science graduate fellowship program grant DE-FG02-97ER25308 and
used resources of the National Energy Research Scientific
Computing Center, which is supported by the Office of Science of
the U.S. Department of Energy under Contract No.
DE-AC03-76SF00098.
\end{acknowledgments}

{}

\end{document}